\documentclass[12pt]{article}
\newcommand{\nn}{\nonumber}
\newcommand{\lb}{\label}

\newcommand{\qq}{\qquad}

\newcommand{\eq}{\equiv}

\newcommand{\be}{\begin{equation}}
\newcommand{\ee}{\end{equation}}
\newcommand{\ba}{\begin{eqnarray}}
\newcommand{\ea}{\end{eqnarray}}
\renewcommand{\a}{\alpha}
\renewcommand{\b}{\beta}

\renewcommand{\k}{\xi}
\renewcommand{\d}{\partial}

\newcommand{\s}{\sigma}
\newcommand{\de}{\delta}
\newcommand{\R}{\Rightarrow}

\newcommand{\const}{{\rm const}}

\newcommand{\ga}{\gamma}

\renewcommand{\SS}{{\cal S}}
\newcommand{\HH}{{\cal H}}
\newcommand{\LL}{{\cal L}}
\newcommand{\NN}{{\cal N}}
\newcommand{\KK}{{\cal K}}
\newcommand{\DD}{{\cal D}}
\newcommand{\OO}{{\cal O}}
\renewcommand{\AA}{{\cal A}}

\renewcommand{\L}{{\bf L}}
\renewcommand{\H}{{\bf H}}
\newcommand{\T}{{\bf T}}
\newcommand{\M}{{\bf M}}
\newcommand{\I}{{\bf I}}
\newcommand{\D}{{\bf D}}
\newcommand{\A}{{\bf A}}
\renewcommand{\R}{{\bf R}}
\newcommand{\diag}{{\rm diag}}
\newcommand{\z}{\zeta}
\newcommand{\w}{\omega}
\newcommand{\ul}{\underline}

\newcommand{\Ts}{{\rm Ts}}

\begin{document}
\begin{center}
{\Large \bf Equivalence of the super Lax and local Dunkl
operators for Calogero-like models\footnote{
Accepted for publication in: Journal of Physics A: Mathematical and General, URL: www.iop.org.
}.
}\\
\end{center}
\vspace{1cm}
 A. I. Neelov\\ \\
{\small 
Institute of Physics, University of Basel, Klingelbergstrasse 82,
CH-4056 Basel, Switzerland
and\\ Department of Theoretical Physics, University of
Sankt-Pe\-ters\-burg, 198504 Sankt-Pe\-ters\-burg,
Rus\-sia.\\ \\
E-mail: alexey.neelov@unibas.ch \\ \\ \\
{\bf Abstract.\ \ } Following Shastry and Sutherland I construct
the super Lax operators for the Calogero model in the oscillator
potential. These operators can be used for the derivation of the
eigenfunctions and integrals of motion of the Calogero model and
its supersymmetric version. They allow to infer several relations
involving the Lax matrices for this model in a fast way. It is
shown that the super Lax operators for the Calogero and Sutherland
models can be expressed in terms of the supercharges and so called local Dunkl operators
constructed in our recent paper with M. Ioffe. 
Several important relations involving Lax matrices and Hamiltonians of the Calogero and Sutherland
models are easily derived from the properties of Dunkl operators.
}

\section*{\large\bf \quad 1. Introduction}

The most well-known exactly solvable and integrable quantum
systems of $N$
 particles on a line are given in \cite{Perelomov},
\cite{genlax}. One of them is the Calogero model \cite{Calogero1}-
\cite{Ruhl}, with the Hamiltonian:
\ba
  H=-\Delta +\w^2\sum_{i=1}^N x_i^2 + \sum_{i\ne j}^N {l^2-l\over
(x_i-x_j)^2}.\label{cal0}
\ea
When  $\w=0$ this model is called the free Calogero or
Calogero-Moser \cite{Moser} one (following the notations of
\cite{WU}).

Another is the trigonometric Sutherland or TS model
\cite{sinh}-\cite{Lap0} with the Hamiltonian
\ba
H=-\Delta+\sum_{i\ne j}^N{l^2-l\over\sin^2(x_i-x_j)}.\label{suth0}
\ea
There is also a hyperbolic variant of the Sutherland model (HS)
\cite{sinh} where there is a hyperbolic sinus in the denominator.
For brevity we will call the three models above the Calogero-like
ones.

These models correspond to the $A_{N-1}$ root system; generalizations
for other root systems also exist
\cite{Perelomov},\cite{Sasakibig}-\cite{Turbiner}.

The formalism of quantum Lax operators \cite{Perelomov},
\cite{genlax},
  \cite{WU}, \cite{sinh},\cite{UHW}-\cite{BMS}, \cite{Sasakibig}
 plays an important role in the proof of the
integrability of the Calogero-like models and derivation of their
eigenfunctions.

The supersymmetric \cite{Rittenberg},\cite{abi} generalization of
 the Calogero model was constructed in \cite{Freedman}, \cite{Vasiliev1}, \cite{Turbiner} and that of the Sutherland model was considered
in \cite{Shastry}, \cite{Lap}.


In the paper \cite{Shastry} the super Lax operators were set
forth. These operators are bilinears in the fermionic variables,
the coefficients being the standard quantum Lax matrices. The
super Lax operators allow one to derive the standard relations
involving the Lax matrices in a faster and simpler way.

Apart from the Lax formalism  there is another,
powerful approach to the proof of the
 integrability and exact solvability  of the Calogero-like models
 that
uses the Dunkl operators \cite{dunkl}-\cite{Sasin}. Its
supersymmetric generalization was constructed in \cite{Lap},
\cite{Des2}, \cite{Des3}. In \cite{inter} another relation between
the Dunkl operators and supersymmetry was considered. Namely, the
so called local Dunkl operators were constructed that intertwined
the matrix Calogero-like Hamiltonians corresponding to some
irreducible representations of $S_N$. For the class of Young
diagrams described in \cite{mag} some local Dunkl operators were
found to coincide with the components of the supercharges (after
the separation of the center of mass (CM) part in the latter). This derivation is
analogous to the projection method of \cite{Lap} that works only
for the supersymmetric models.

The main result of the present work is that another class of the
local Dunkl operators of \cite{inter} coincides with the
CM-independent part of the components of the super Lax operators
of \cite{Shastry}.

Thus one has a means to construct the Lax operators for a given system provided
it possesses a set of Dunkl operators. This can be useful e.g. for the Calogero-like systems
for the root systems other than $A_{N-1}$ or for elliptic Calogero models.

The paper is organized as follows:

In Section 2 we briefly review the formalism of the
supersymmetric quantum mechanics (SUSY QM) \cite{Rittenberg},
\cite{abi} and its application to the Calogero-like models. The
super Lax operators for the free models \cite{Shastry} are
constructed. The components of these operators in the
one-fermionic sector turn out to coincide with the usual Lax
matrices.  We also construct
the super Lax operator for the Calogero model which we believe is
new. It can be used for the construction of the eigenstates of the
model and for the proof of its integrability. Some useful
identities for the total sums of the Lax matrices \cite{Sasakibig}
are formulated. They are to be proved in the subsequent sections.

In Section 3 the bosonic \cite{Reed} and fermionic Jacobi
variables with reference to the Calogero-like models \cite{mag},
\cite{eft}. are introduced. The separation of the CM part in the superhamiltonian
and supercharges \cite{mag} is briefly reviewed.
It is shown that in the case of the Calogero model one can obtain the
identities for the total sums of Lax matrices given in \cite{UHW},
\cite{UWH} from the properties of the super Lax operators constructed in Section 2. 

In the beginning of Section 4 the local Dunkl operators
\cite{inter} are presented. The relations in which they intertwine
the matrix Hamiltonians for the Calogero-like models are given. 

Then a special kind of the Clebsh-Gordan coefficients for the
local Dunkl operator of a free Calogero-like model is constructed
with the help of fermionic variables. Thus we give an explicit
example of the exactly solvable Dirac-like operator of
\cite{inter}. The new local Dunkl operator can be viewed as a
component of a certain super Lax-like operator, bilinear in
fermions. This super Lax-like operator turns out to coincide with
the CM-independent part of the usual super Lax operator \cite{Shastry} written in
the Jacobi variables. Therefore one can infer the fact that the
super Lax operator commutes with the superhamiltonian from the
intertwining relations of the local Dunkl operators and matrix
Hamiltonians derived in \cite{inter}. The CM-dependent part of the
super Lax operator is expressed in
terms of the supercharge operators, which allows us to prove an
identity from
 \cite{UHW}, \cite{UWH}.

Then we use the same Clebsh-Gordan coefficients for the local
Dunkl operators for the Calogero model. The result again has the
form of components of certain super Lax-like operators. The
latter, instead of commuting with the superhamiltonian, will obey
oscillator-like commutation relations with it. As in the free
case, the new super Lax-like operators coincide with the
CM-independent components of the usual super Lax operators written
in the Jacobi variables. This again allows one to infer the
oscillator-like commutation relations between the super Lax
operators and the Hamiltonian from the intertwining relations of
the local Dunkl operators and matrix Hamiltonians. For the
Calogero model the CM-dependent part of the super Lax operator is
again expressed in terms of the
supercharge operators, which allows us to prove an identity from
\cite{WU}.

The possible extension of the results of the paper onto the case
of the Calogero-like models corresponding to general root systems
is briefly discussed in the last subsection.

\section*{\large\bf 2.\quad Supersymmetric Calogero-like models.}
\subsection*{\it 2.1. \quad Multidimensional SUSY QM \cite{abi}.}

 The
supersymmetric quantum system for arbitrary number of dimensions
$N$ consists \cite{abi}  of the superhamiltonian $\HH$
and the
supercharges\footnote { Here and below the indices $i,j,k,\ldots $
range from 1 to $N $. }:
\ba
Q^-\equiv
 \sum_{j=1}^N \psi_{j} Q^+_j \qquad
Q^+=(Q^-)^\dagger= \sum_{j=1}^N \psi_{j}^+ Q^-_j 
 \label{Qpm}
\ea
 with the algebra
\ba
(Q^+)^2=(Q^-)^2=0\qq \HH = \{Q^+,Q^-  \} \label{HQQ}
\ea
\be
[\HH,Q^\pm]=0 \label{sus}
\ee
  where $\psi_i,\ \psi_i^+=(\psi_i)^\dagger$ are fermionic operators:
\ba
\{\psi_i,\psi_j\}=0\qquad \{\psi_i^+,\psi_j^+\}=0 \qquad
\{\psi_i,\psi_j^+\}=\delta_{ij}.\label{ant}
\ea

The Hamiltonian and supercharge operators act in the tensor product of the fermionic Fock space
with the basis
\ba
 \psi_{i_1}^+\ldots\psi_{i_M}^+|0>&\equiv& |i_1\ldots i_M>\qquad i_1<...<i_M\leq N\qquad M\leq N\label{bpsd}\\
   \psi_{i}|0>&=&0\qquad i\leq N  \nonumber
\ea
and some bosonic Fock space where the operators $Q_i^\pm$ act. From this moment on we will not mention the 
bosonic Fock space for brevity.

The superhamiltonians condidered in this text conserve the
fermionic number $\NN\equiv \sum_{j=1}^N\psi_j^+\psi_j$. Hence,
they have the following block-diagonal form in the basis
(\ref{bpsd}):
\ba
  \HH = \diag(H^{(0)},\H^{(1)}...,\H^{(N-1)},H^{(N)})\label{Hdiag}
 \ea
where the matrix operator $ \H^{(M)}$ with dimension\footnote{The $C_{N}^{M}$ here are the binomial coefficients.} $
C_{N}^{M}\times C_{N}^{M}$ is the component of $\HH$ in the
subspace with fixed fermionic number $M$. The components with $M$
equal to 0 and $N$ are thus scalar operators, and are not marked
by boldface.

\subsection*{\it 2.2. \quad Supersymmetric Calogero-like models \cite{Freedman}-\cite{mag}.}

The free supersymmetric Calogero-like models are characterized by
the bosonic parts of the supercharges (\ref{Qpm}) of the form
\ba
Q_l^{\pm}=\mp\partial_l-\sum_{l \neq k}^NV(x_l-x_k)\equiv
\mp\partial_l-\sum_{l \neq k}^NV_{lk} \label{Qf}
\ea
 where V(x) are given in the Table and $V_{lk}\equiv V(x_l-x_k)$.
\begin{table}
{Table}\\
\begin{tabular}{ccccc}
\hline \hline {\hskip 0.9 cm} Name of {\hskip 0.9 cm}  & {\hskip 0.8 cm} $V(x)$ {\hskip
0.8 cm} & {\hskip 0.8 cm} $V'(x)$ {\hskip 0.8 cm} &{\hskip 0.11 cm} $E_0$\ {\hskip 0.11 cm} \\
model
\\
\hline TS &  $l\cot x$ & $-l/\sin^2x$&
 $-(N-2)(N-1)N{l^2/ 3}$ \\ \\
HS &  $l\coth x$ & $-l/\sinh^2x$&
 $-(N-2)(N-1)N{l^2/ 3}$ \\ \\
Free Calogero &  $l/x$ &
 $-l/x^2$ & $0$ \\
\\
\hline \hline
\end{tabular}
\end{table}

With such supercharges the superhamiltonian (\ref{HQQ}) turns
into \cite{mag}
\ba
  \HH=-\Delta +\sum_{i\ne l}^NV_{il}^2+
  \sum_{i\ne j}^N \KK_{ij}V_{ij}'-E_0.  \label{HH}
\ea
The constants $E_0$ are
given in the Table. The operator $\KK_{ij}$ \cite{Shastry} has the
form
\ba
  \KK_{ij}\equiv \psi^+_i\psi_j+\psi^+_j\psi_i-\psi^+_i\psi_i-
\psi^+_j\psi_j+1=
1-(\psi_i^+-\psi_j^+)(\psi_i-\psi_j)=\nonumber\\
 = \KK_{ji}=( \KK_{ij})^\dagger
\label{Kij}
\ea
and is the fermionic exchange operator:
\ba
 \KK_{ij}\psi^+_i&=& \psi^+_j \KK_{ij}\qquad
 \KK_{ij}\psi_i= \psi_j \KK_{ij}\label{pe1} \\
 \KK_{ij}\psi^+_k&=&\psi^+_k  \KK_{ij}\qquad
\KK_{ij}\psi_k=\psi_k  \KK_{ij}\quad k\ne i,j. \label{pr2}
\ea

The Calogero model is characterized by 
the bosonic parts of the supercharges (\ref{Qpm}) of the form
\ba
Q_l^{\pm}=\mp\partial_l+\w x_l-l\sum_{l \neq k}^N(x_l-x_k)^{-1}\lb{bocal}.
\nn
\ea
Accordingly, the superhamiltonian (\ref{HQQ}) of the model has the
form
\ba
  \HH=-\Delta +\w^2\sum_{i} x_i^2 + \sum_{i\ne j} {l^2-l\KK_{ij}\over
(x_i-x_j)^2}+2\w\NN-  \w\biggl(1+(N-1)(Nl+1)\biggr).\label{HHCO}
\ea

The exchange operator $\KK_{ij}$ in (\ref{HH}), (\ref{HHCO})
commutes with $\NN$, and therefore assumes a block-diagonal form
in the basis (\ref{bpsd}), similarly to the superhamiltonian:
\ba
  \KK_{ij} = \diag(T_{ij}^{(0)},\T_{ij}^{(1)}...,\T_{ij}^{(N-1)},T_{ij}^{(N)}).\label{Kdiag}
\ea
The components (\ref{Hdiag}) of the superhamiltonian have the form
\ba
  \H^{(M)}=\biggl[-\Delta +\sum_{i\ne l}^NV_{il}^2-E_0\biggr]\I+
  \sum_{i\ne j}^N \T_{ij}^{(M)}V_{ij}'  \nn
\ea
for the free Calogero-like models, and
\ba
  \H^{(M)}=
 \biggl[-\Delta
+\w^2\sum_{i}^N x_i^2 +\w\biggl(2M- 1
-(N-1)(Nl+1)\biggr)\biggr]\I+\sum_{i\ne j}^N
{l^2\I-l\T_{ij}^{(M)}\over (x_i-x_j)^2}\label{hmc}
\ea
for the Calogero model.

One can easily see that $T_{ij}^{(0)}=1=-T_{ij}^{(N)}$. Thus the
component $H^{(0)}$ coincides up to an additive constant with the
scalar hamiltonian (\ref{suth0}) for the Sutherland model and with
(\ref{cal0}) for the Calogero model.

The elements of the matrix $\T_{ij}^{(1)}$ have the form
\ba
(T_{ij}^{(1)})_{lk}\equiv
\delta_{lk}-\delta_{li}\delta_{ki}-\delta_{lj}\delta_{kj}
+\delta_{li}\delta_{kj}+\delta_{lj}\delta_{ki}.\label{Tij1}
\ea

\subsection*{\it 2.3. \quad The super Lax operators.}

As noted in \cite{Shastry} , the superhamiltonian (\ref{HH})
satisfies the following commutation relation:
\ba
[\HH,\LL]=0\label{LLH}
\ea
where the operator $\LL$ is the so called super Lax operator given
by
\ba
\LL=L_{km}\psi_k^+\psi_m\qquad
L_{km}=-i\d_k\de_{km}+i(1-\de_{km})V_{km}.\label{LL}
\ea
Here, $L_{km}$ are the elements of the well-known Lax matrix $\L$,
and $\psi_k^+, \psi_m$ are the fermionic operators (\ref{ant}).

In the
section 4.2 of this paper we present an alternative proof of
(\ref{LLH}) using the Dunkl operators.

One may also note that $[\NN,\LL]=0$, so the super Lax operator
conserves the fermionic number and has the block-diagonal form:
\ba
  \LL = \diag(0,\L^{(1)}...,\L^{(N-1)},L^{(N)}).\label{diag}
\ea
Note that $L^{(N)}=-i\sum_k\d_k$.

We will use below the following consequence of the anticommutaion relations (\ref{ant}): For a fermionic quantity
\ba
\AA=\sum_{k,l}A_{kl}\psi^+_k\psi_l\label{AAbil}
\ea
the matrix elements in the
one-fermionic sector are
\ba
<i|\AA|j>=A_{kl}<0|\psi_i\psi_k^+\psi_l\psi_j^+|0>=A_{ij}\label{Aij}
\ea
so its first block on the diagonal in the form (\ref{diag}) is
$\A^{(1)}=\A$. For example, $\L^{(1)}=\L$.

The standard relation involving the Lax matrices is:
\ba
[\L,H^{(0)}]=[\M,\L]\label{lhm}
\ea
where $H^{(0)}$ is the Hamiltonian of a scalar free Calogero-like model,
and the elements of $\M$ have the form:
\ba
M_{lk}=2(1-\de_{lk})V_{lk}'-2\de_{lk}\sum_{j\ne k}V_{kj}'.\label{Mlk}
\ea
Eq. (\ref{lhm}) was shown in \cite{Shastry} to follow from (\ref{LLH}), but not vice versa.

The Lax matrix for the free Calogero-like models satisfies the
following identity \cite{WU0}:
\ba
\Ts(\L^2)=H^{(0)}\label{TSL}
\ea
which is used in the proof of integrability of the free
Calogero-like models \cite{BMS}. For a matrix $\A$ the total sum
$\Ts$ is defined as
\ba
\Ts \A=\sum_{i,j=1}^NA_{ij}.\nonumber
\ea
Eq. (\ref{TSL}) will also be proven in Section 4.2.

The following identity is also true \cite{Shastry}:
\ba
[H^{(0)},I_n]=0\qquad I_n=\Ts\L^n.\label{HIn}
\ea
 The involution of the quantities $I_n$ is
proved in \cite{UHW}, \cite{UWH}.

It turns out that the construction of the super Lax operators is
possible for the Calogero model too. To the author's knowledge this construction has not been proposed before; thus the rest of the Subsection contains new material.
Namely, define the following fermionic operator:
\ba
 \LL^\pm=L^\pm_{km}\psi_k^+\psi_m
\qquad L^\pm_{km}=L_{km}\pm i\w x_k\de_{km}\qquad
L_{km}=-i\d_k\de_{km}+i{1-\de_{km}\over x_k-x_m}  \label{Lpm}
\ea
where $L_{km}$ are the elements of the Lax matrix  for the free
Calogero model. The operators (\ref{Lpm}) and the superhamiltonian
(\ref{HHCO}) satisfy the following generalization of (\ref{LLH}):
\ba
[\HH,\LL^\pm]=\pm 2\omega  \LL^\pm.\label{HLLpm}
\ea
 The
proof of these relations can be found in the Subsection 4.3. Eq.
(\ref{HLLpm}) describes an oscillator-like algebra and hence can
be used for the construction of the spectrum of the
superhamiltonian (\ref{HHCO}) and proof of its integrability.
Namely, the ground
state wave function for the (super)Calogero Hamiltonian (\ref{cal0}),(\ref{HHCO}) is 
\cite{Calogero1},\cite{Freedman}:
\ba
\psi_0=\exp\biggl(-{\w\over 2}\sum_{j=1}^Nx_j^2\biggr)\prod_{i <
k}^N|x_i-x_j|^l.\label{psiCO}
\ea
Applying powers of the operators (\ref{Qpm}), (\ref{Lpm}) to this
wave function one can get the excited states of $\HH$. The
integrals of $\HH$ are linear combinations of the monomials in
$Q^\pm$, $\LL^\pm$ in which the power of $\LL^+$ is equal to the
power of $\LL^-$. Examples of such are
\ba
\LL_1=\LL^+\LL^-\qquad \LL_2=\LL^-\LL^+.\label{LLk}
\ea

Similarly to the free Calogero models, $[\NN,\LL^\pm]=0$, so
\ba
  \LL^\pm = \diag(0,\L^{(1)\pm}...,\L^{(N-1)\pm},L^{(N)\pm}).\nonumber
\ea
It follows from (\ref{Aij}) that $\L^{(1)^\pm}=\L^\pm$ where
$\L^\pm$ is the matrix with the elements (\ref{Lpm}).

One can infer the usual relations \cite{WU0} involving the Lax matrices from
(\ref{HLLpm}) in the following way:
\ba
\pm 2\w\L^\pm=[\L^\pm,\H^{(1)}]=[\L^\pm,\H^{(1)}-\I H^{(0)}+\I
H^{(0)}].\nonumber
\ea
where $\H^{(1)}$ is the first component (\ref{hmc}) of the superhamiltonian.
Hence,
\ba
[H^{(0)},\L^\pm]=[\L^\pm,\M]\pm 2\w\L^\pm\nonumber
\ea
where $\M\equiv \H^{(1)}-\I \biggl(H^{(0)}+2\w\biggr)$ is the same standard
matrix (\ref{Mlk}) as in the free case. It follows from (\ref{hmc}),(\ref{Tij1})
that its elements are
\ba
M_{mk}=2l(\de_{mk}-1)(x_m-x_k)^{-2}+2l\de_{mk}\sum_{j\ne
k}(x_k-x_j)^{-1}.\nonumber
\ea

For the Calogero model we can derive an analog of (\ref{TSL}).
Namely, define the quantities
\ba
\L_1=\L^+\L^-\qquad \L_2=\L^-\L^+.\nonumber
\ea
The matrices $\L_j$ are the components of the operators $\LL_j$
(\ref{LLk}) in the sector $\NN=1$. It turns out \cite{WU0} that
\ba
H^{(0)}=\Ts\L_1=\Ts\L_2+\const.\label{TsL1}
\ea
A variant of proof can be found in Subsection 4.3. 

An analog of (\ref{HIn}) can also be proved:
\ba
[H^{(0)},I_{jn}]=0 \qquad I_{jn}=\Ts\L_j^n\qquad j=1,2\label{HIjn}
\ea
a proof is given in Section 3. The involution of the quantities
$I_{jn}$ is proved in \cite{WU}, \cite{Sasakibig}.

One can also define the following operators \cite{WU0}:
\ba
O_p^m=\Ts\biggl((\L^-)^m(\L^+)^p\biggr)\label{Omp}
\ea
that commute with  $H^{(0)}$ as
\ba
[H^{(0)},O_p^m]=2(p-m)\w O_p^m.\label{HO}
\ea
The proof is again given in Section 3. Applying the operators
(\ref{Omp}) to the ground state wave function (\ref{psiCO}) one
gets the excited states of $H^{(0)}$ for the Calogero model.

\section*{\large\bf 3.\quad The Jacobi variables and SUSY QM.}
\subsection*{\it 3.1. \quad Definitions \cite{mag}.}

The bosonic \cite{Reed} and fermionic \cite{mag} Jacobi variables are defined as
\ba
y_k=R_{km}x_m\qq \phi_k=R_{kl}\psi_l \label{jac}
\ea
where $R_{kl}$ is a real orthogonal matrix; see \cite{mag} for details. 
In this text it will be important for us that $R_{Nl}=N^{-1/2}$, i.e.,
\ba
y_N = \frac{1}{\sqrt{N}}\sum_{i=1}^N x_i\qq
\phi_N = \frac{1}{\sqrt{N}}\sum_{i=1}^N \psi_i.\nn
\ea
The new fermionic variables (\ref{jac}) satisfy the standard anticommutation
relations:
\ba
\{\phi_k,\phi_l\}=0\qquad \{\phi_k^+,\phi_l^+\}=0 \qquad
\{\phi_k,\phi_l^+\}=\delta_{kl}.\label{phiant}
\ea

With the help of the fermionic Jacobi variables one can separate
the center of mass term in the supercharges (\ref{Qpm}) in the
following way \cite{mag}:
\ba
Q^\pm= q^\pm+Q_C^\pm\qq \HH=h+H_C\label{Qqpm}
\ea
where
\ba
Q_C^{-}\equiv -\phi_N\frac{\partial}{\partial y_N}\qquad
Q_C^+=\phi_N^+\frac{\partial}{\partial
y_N}\qq
H_C=-\d^2/\d y_N^2\label{QC}
\ea
for the free models, and
\ba
Q_C^-&\equiv& \phi_NQ_N^+\qq Q_C^+\equiv \phi_N^+Q_N^-\qq
Q_N^\pm=\mp {\d\over\d y_N}+\w y_N\label{qccal}\\
H_C&=&-d^2/dy_N^2+\w^2y_N^2+\w(2\phi^+_N\phi_N-1)\label{hccal}
\ea
for the Calogero model.

These new quantities satisfy the relations of the following superalgebra
\cite{mag}:
\ba
(q^\pm)^2&=&(Q_C^\pm)^2=\{q^\pm,Q_C^\pm\}=0\nn\\
\{q^+,q^-\}&=&h\qquad
\{Q^+_C,Q^-_C\}=H_C\qq [h,H_C]=0 \label{HHh}\\
\bigl[\HH,q^\pm\bigr]&=&[h,q^\pm]=[H_C,q^\pm]=
\bigl[\HH,{Q_C}^{\pm} \bigr]=[h,{Q_C}^{\pm}]=[H_C,{Q_C}^{\pm}]=0.\nn
\ea

\subsection*{\it 3.1. \quad Application to the Lax operators.}

If one uses the fermionic Jacobi variables, it is natural to go from the basis (\ref{bpsd}) to a new
one\footnote{The indices of the Jacobi variables denoted by Greek
letters range from 1 to N-1; those denoted by Latin letters range
from 1 to N} (here we follow \cite{mag}):
\ba
  \phi_{\b_1}^+\ldots\phi_{\b_M}^+ |0>&\equiv& |\b_1...\b_M>\equiv |\b>\qquad
\phi_N^+\phi_{\b_1}^+\ldots\phi_{\b_M}^+ |0>\equiv
|N\b>\label{ste}\\
  \b_1&<&...<\b_M\qquad M<N.\nonumber
\ea
In the new basis (\ref{ste}) , the superhamiltonian (\ref{HH}) of the free
Calogero-like models will take the form:
\ba
\HH=\diag(\widetilde{H}^{(0)},\widetilde{
\H}^{(1)},\ldots,\widetilde{ \H}^{(N-2)},\widetilde{ H}^{(N-1)},
\widetilde{ H}^{(0)},\widetilde{ \H}^{(1)},\ldots,\widetilde{
\H}^{(N-2)},\widetilde{ H}^{(N-1)})\nn
\ea
where
\ba
  \widetilde\H^{(M)}=\biggl[-\Delta +\sum_{i\ne l}^NV_{il}^2-E_0\biggr]\I+
  \sum_{i\ne j}^N \widetilde{ \T}_{ij}^{(M)}V_{ij}' \lb{tilh}
\ea
and ${\widetilde \T}_{ij}^{(M)}$ are matrices\footnote{One
should not confuse ${\widetilde \T}_{ij}^{(M)}$ with
$\T_{ij}^{(M)}$ from (\ref{Kdiag}) which corresponds to a {\it
reducible} representation of $S_N$.} with the elements
\ba
({\widetilde T}_{ij}^{(M)})_{\ga \b}=<\ga_M\ldots\ga_1|\KK_{ij}|\b_1\ldots\b_M>\label{defT}
\ea
where $\KK_{ij}$ is the fermionic exchange operator (\ref{Kij}).
It is proved in \cite{mag} that such matrices form the
representation\footnote{We will denote the irreducible
representations of $S_N$ by their Young diagrams. The standard
notation \cite{Ham} for the Young diagram containing $\lambda_i$
cells in the $i$-th line is $(\lambda_1,\ldots,\lambda_n)$; if a
diagram contains $m$ identical lines with $\mu$ cells, it is
denoted by $(\ldots,\mu^m,\ldots)$. } of $S_N$ with the Young
diagram $(N-M,1^M)$.

The superhamiltonian (\ref{HHCO}) of the Calogero model in the
Jacobi basis will have the form:
\ba
\HH&=&\diag(\widetilde{H}^{(0)},\widetilde{
\H}^{(1)},\ldots,\widetilde{ \H}^{(N-2)},\widetilde{ H}^{(N-1)},
\widetilde{ H}^{(0)}+2\w,\widetilde{ \H}^{(1)}+\nonumber\\
&+&2\I\w,\ldots,\widetilde{ \H}^{(N-2)}+2\I\w,\widetilde{
H}^{(N-1)}+2\w)\lb{tilhc}
\ea
where
\ba
  \widetilde\H^{(M)}=
 \biggl[-\Delta
+\w^2\sum_{i}^N x_i^2 +\w\biggl(2M- 1
-(N-1)(Nl+1)\biggr)\biggr]\I+\sum_{i\ne j}^N
{l^2\I-l\widetilde\T_{ij}^{(M)}\over (x_i-x_j)^2}.\nonumber
\ea


Now one can use the above formalism from (\cite{mag}) to prove (\ref{HO})
in the same  way as (\ref{HIn}) was proved in \cite{Shastry}.

First
take into account that for any fermionic quantity $\AA$ that
commutes with $\NN$,
\ba
\Ts\A^{(1)}=\sum_{i,j=1}^NA^{(1)}_{ij}=\sum_{i,j}<i|\AA|j>=
 N<N|\AA|N>\label{TsA}
\ea
where $\A^{(1)}$ is the component of $\AA$ in the sector with
$\NN=1$. Then, define the quantities
\ba
\OO^m_p=(\LL^-)^m(\LL^+)^p.\nonumber
\ea
It follows from (\ref{HLLpm}) that
\ba
[\HH,\OO^m_p]=2(p-m)\omega\OO^m_p.\nonumber
\ea
From (\ref{TsA}) we get:
\ba
O^m_p=\Ts\bigl[(\L^-)^m(\L^+)^p\bigr]=N<N|\OO^m_p|N>.\nonumber
\ea
Therefore,
\ba
&&2(p-m)\omega O^m_p=2(p-m)\omega N<N|\OO^m_p|N>=N<N|[\HH,\OO^m_p]|N>=\nonumber\\
&&=N<N|\HH\OO^m_p|N>-N<N|\OO^m_p\HH|N>=(H^{(0)}+2\w)N<N|\OO^m_p|N>-\nonumber\\ 
&&-N<N|\OO^m_p|N>(H^{(0)}+2\w)=[H^{(0)},O^m_p].\nonumber
\ea
In exactly the same way one can deduce (\ref{HIjn}) from (\ref{LLH}).

\section*{\large\bf 4.\quad Connection between the local Dunkl
operators and the super Lax operators.}
\subsection*{\it 4.1. \quad Intertwining relations involving the
local Dunkl operators.}

In a recent paper \cite{inter} we considered the matrix
Calogero-like Hamiltonians of the form
\ba
  \H^{A}=\biggl[-\Delta +\sum_{i\ne l}^NV_{il}^2\biggr]\I+
  \sum_{i\ne j}^N \T_{ij}^{A}V_{ij}'  \label{HA}
\ea
where $A$ is an irreducible representation of the group $S_N$ of
permutations of $N$ particles, and $\T_{ij}^{A}$ are the matrices
of this representation.

We will need below the representation $L$ with the Young diagram
$(N-1,1)$. Suppose we have an irreducible representation $A$ of $S_N$,
 such that the interior product
$L\times A$ contains $A$. Then   the following commutation relation
is true \cite{inter} :
\ba
[\H^A,\D^{AA}]=0\label{HD}
\ea
where $\D^{AA}$ is the so called local Dunkl operator. It is a
$\dim A\times\dim A$ matrix with elements
\ba
 D^{AA}_{\s\a}=(\k\b|\s)R_{\k k}
\biggl[-i\d_k\de_{\b\a} + i \sum_{m \neq k}
V_{km}(T_{km}^A)_{\b\a}\biggr]. \label{DBA}
\ea
Here, $R_{\k k}$ is the matrix of transition from the particle
coordinates to the Jacobi ones; $(\k\b|\s)\equiv(L\,\k\ ,\
A\,\b\,|\,A\,\s)$ are the Clebsh-Gordan coefficients for the
contribution of $A$ in $L\times A$.

Note that the SUSY QM intertwining relations for the Calogero-like systems 
can also be deduced from the local Dunkl operators \cite{inter}.

It was proved in \cite{mag} that for the TS model, Eq. (\ref{HD})
allows us to find the spectrum of $\D^{AA}\equiv \D^A$ because we
know the spectrum of $\H^A$. However, the definition (\ref{DBA})
of $\D^A$ contains a Clebsh-Gordan coefficient $(L\,\k\ ,\
A\,\b\,|\,A\,\s)$ that is relatively hard to find, except for the
cases discussed below and in \cite{inter}.

In the case of the Calogero model the
following analog of (\ref{HD}) was set forth in \cite{inter}:
\ba
[\H^A,\D^{AA\pm}]=\pm 2\w\D^{AA\pm} \label{comw}
\ea
where $\H^A$ is the matrix Calogero hamiltonian for the
representation $A$:
\ba
\H^A=\biggl[ -\Delta + \omega ^2 \sum_i x_i^2 +\sum_{i\ne j}{l^2
\over(x_i-x_j)^2}+N\omega\biggr]\I-\sum_{i\ne j}\biggl({l\over
(x_i-x_j)^2 }+a\biggr)\T_{ij}^A \label{HCOA}
\ea
and the elements of the matrix $\D^{AA\pm}$ are
\ba
D^{AA\pm}_{\s\a}&=&(\k\b|\s)R_{\k j} \biggl[(-i\d_j\pm i\omega
x_j)\de_{\b\a} + il \sum_{m \neq j} {(T_{jm}^A)_{\b\a}\over
x_j-x_m} \biggr].\label{DAw}
\ea

We will see in Section 4.3 that the components of the operators
$\LL^\pm$ can be reduced to a partial case of (\ref{DAw}) (see
Eqs. (\ref{LL+}), (\ref{LL-})).

\subsection*{\it 4.2. \quad The connection between the local Dunkl
operators and the super Lax ones.}

In this Subsection we are to prove that the super Lax
operator (\ref{LL}) can be expressed in terms of the local Dunkl
operators (\ref{DBA}) and the commutation relations (\ref{LLH})
follow from (\ref{HD}).

Let us suppose that 
\ba
A=(N-M,1^M)\lb{anm}
\ea
in (\ref{HD}), (\ref{DBA}).
Then we can define the Clebsh-Gordan coefficients in (\ref{DBA})
in the following way:
\ba
(1\,\xi,M\,\b|M\,\z)=<\z|C_\xi|\b>\qquad C_\xi\equiv R_{\xi
k}\psi_k^+\psi_k=C_\xi^\dagger\label{C}
\ea
where $\psi_i,\psi_i^+$ are the fermionic variables satisfying
 (\ref{ant}); $R_{\xi k}$ is the matrix of transition (\ref{jac}), (\ref{jac}) from the particle
coordinates to the Jacobi ones; $|\b>$ are the states from (the
first half of) the basis (\ref{ste}), such that $\NN |\b>=M|\b>;\
\ \phi_N|\b>=0$.

This is possible because the coefficients (\ref{C}) satisfy the
following characteristic condition of the Clebsh-Gordan
coefficients \cite{Ham}:
\ba
(T_{ij}^L)_{\a\xi} ({\widetilde T}_{ij}^{(M)})_{\ga
\b}(1\,\xi,M\,\b|M\,\z)= (1\,\a,M\,\ga|M\,\nu)({\widetilde
T}_{ij}^{(M)})_{\nu\z}\label{TTC}
\ea
where ${\widetilde \T}_{ij}^{(M)}$ is the matrix (\ref{defT}), and
\ba
\T_{ij}^L={\widetilde \T}_{ij}^{(1)}\label{til}
\ea
is a matrix from the representation $L$. 
The proof of (\ref{TTC}) can be found in Appendix 2.

As shown in \cite{mag}, for the representaions from the class (\ref{anm})
one can go from the matrix Hamiltonian $\H^A$ (\ref{HA}) in (\ref{DBA})  to
${\widetilde \H}^{(M)}=\H^{A}-E_0\I$ that is given by (\ref{tilh}), $E_0$ being
given in the Table.

Now we can plug the Clebsh-Gordan coefficients (\ref{C}) into the
definition (\ref{DBA}). After some algebra (see Appendix 3 for
details) we arrive at the equality
\ba
 D^{A}_{\s\a}&\equiv& D^{(M)}_{\s\a} =<\s|C_\xi|\b>R_{\k k}\biggl[-i\d_k\de_{\b\a} +
  i \sum_{m \neq k}
V_{km}({\widetilde T}_{km}^{(M)})_{\b\a}\biggr]=\nonumber\\
&=&<\s|\LL|\a> +iN^{-1/2}M{\d\over\d y_N}\de_{\s\a}\label{DL}
\ea
where $\LL$ is the super Lax operator (\ref{LL}). Thus we see that
the matrix elements of the local Dunkl operator in the basis
(\ref{ste}) coincide with the matrix elements of the super Lax
operator, up to a scalar term.

The operator (\ref{DL}) has the structure of a matrix element connecting two
fermionic basis states $<\a|$ and $|\b>$. It is natural to consider a
fernionic operator built from these matrix elements: 
\ba
\DD=\sum_{M,\s,\a}D_{\s\a}^{(M)}\biggl[|\s><\a|+|N\s><\a N|\biggr].\label{DDD}
\ea
In (\ref{DDD}) and all formulae below, the states $|\s>,|\a>$ have fermionic number $M$, if not specified otherwise.

It follows from (\ref{HD}) that the operator (\ref{DDD}) commutes with the superhamiltonian
(\ref{HH}).

The components $\D^{(M)}$ (\ref{DL}) of $\DD$ have smaller dimension
than $\L^{(M)}$, i.e., the block-diagonal structure of $\DD$ is
more detailed than that of $\LL$. Note that $D^{(0)}=0$.

After a couple of pages of calculations we can conclude that
\ba
\DD=\LL+iN^{-1/2}\biggl[Q^+\phi_N-\phi_N^+Q^-+\biggl(\NN-
2\phi_N^+\phi_N\biggr){\d\over\d y_N}\biggr].\label{LLDD}
\ea
The details are given in the Appendix 4.

Eq. (\ref{LLDD}) gives a simple form of the operator $\DD$ and its components $\D^{(M)}$
that can be considered as exactly solvable Dirac-like operators.

It immediately follows from (\ref{LLDD}) that $[\LL,\HH]=0$ since all other operators in 
(\ref{LLDD}) have already been seen to commute with $\HH$. The only nontrivial 
commutaion relation of this kind:
$[\phi_N,\HH]=[\phi_N^+,\HH]=0$
follows from (\ref{Qqpm}),(\ref{QC}).

One can also check that
\ba
[h,\DD]=0\label{HDD}
\ea
where $h=\HH+{\d^2\over\d y_N^2}$
is the center-of-mass independent part (\ref{Qqpm}) of the
superhamiltonian.

Eq. (\ref{HDD}) means that $\DD$ plays the same role for $h$ as
$\LL$ does for $\HH$. However, $h$ and $\DD$ do not depend on the
CM variables\footnote{
In case of $\DD$ it can be proved by rewriting the operator (\ref{LL}) in the Yacobi
variables and using (\ref{LD}).
} $y_N,\ \phi_N,\ \phi_N^+$. Thus we have obtained a
separation of variables in the (super) Lax operators.

One can also go from the Dunkl operators to the Lax ones by using the approach \cite{Lap}
that does not employ Jacobi variables.
However, then it would be difficult to get separation of the center of mass coordinate, and 
obtain the operators $\DD$.


The center of mass terms can also be separated in the supercharges in (\ref{LLDD}), 
according to (\ref{Qqpm}).The result will be
\ba
\LL=\DD+iN^{-1/2}\biggl[\phi_N^+q^--q^+\phi_N- \NN{\d\over\d
y_N}\biggr].\label{LD}
\ea

Eq. (\ref{LD}) can be used for the derivation of (\ref{TSL}):
\ba
\Ts(\L^2)=H^{(0)}.\nonumber
\ea

Namely, taking into account (\ref{TsA}),  we get
\ba
&&\Ts(\L^2)=N<N|\LL^2|N>=N<N|\biggl[
\DD+iN^{-1/2}\biggl(\phi_N^+q^--q^+\phi_N-\nonumber\\&& -
\NN{\d\over\d y_N}\biggr)\biggr]^2|N>=
N<N|iN^{-1/2}\biggl[\phi_N^+q^--\NN{\d\over\d
y_N}\biggr]iN^{-1/2}\biggl[-q^+\phi_N-\nonumber\\&&-\NN{\d\over\d
y_N}\biggr]|N>= -<N|-\phi_N^+q^-q^+\phi_N+\NN^2{\d^2\over\d
y_N^2}|N>=<0|q^-q^++q^+q^--\nonumber\\
&&-{\d^2\over\d y_N^2}|0>=<0|\HH|0>=H^{(0)}\nonumber
\ea
where we have used the fact that $\DD|N>=0;\ <N|\DD=0$, because $D^{(0)}=0$.

We see that (\ref{TSL}) actually follows from the supersymmetry of
the model.

\subsection*{\it 4.3. \quad The super Lax operators for the Calogero model.}

It will be convenient below to rewrite the super Lax operator
(\ref{Lpm}) of the Calogero model in the form:
\ba
\LL^\pm=\LL\pm\de\LL\qquad \de\LL\equiv i\w x_k
\psi_k^+\psi_k\nn
\ea
where $\LL$ is the super Lax operator (\ref{LL}) for the Calogero
model without the harmonic term.

Similarly one can rewrite the local Dunkl operator (\ref{DAw}) as
\ba
\D^{AA\pm}\equiv
 \D^{AA}\pm\de\D^{AA}\nn
\ea
where $\D^{AA}$ is the local Dunkl operator (\ref{DBA}) for the
free Calogero model and $\de\D^{AA}$ is the operator with the
elements
\ba
\de D^{AA}_{\s\a}=(L\,\xi\ ,\ A\,\a\,|\,A\,\s)R_{\xi k}i\w
x_k.\nn
\ea

For the case $A=(N-M,1^M)$ and the choice (\ref{C}) of the
Clebsh-Gordan coefficients we have Eq. (\ref{DL}). Similar 
relation is true for $\de\D^{AA}\equiv\de\D^{(M)}$ and
$\de\LL$:
\ba
\de D^{(M)}_{\ga\de}= <\ga|\de\LL|\de>-i\w N^{-1/2}My_N\de_{\s\a}.\nn
\ea
The proof is completely similar to that of (\ref{DL}), so we omit it.

Similarly to the free case, one can define the operators
\ba
\de\DD&\eq&\sum_{M,\s,\a}\de D_{\s\a}^{(M)}\biggl[|\s><\a|+|N\s><\a N|\biggr]\nn\\
\DD^\pm&\eq&\DD\pm\de \DD=\sum_{M,\s,\a}D_{\s\a}^{(M)\pm}\biggl[|\s><\a|+|N\s><\a N|\biggr]\label{ddpm}
\ea
where $\D^{(M)}$ is the local Dunkl operator for the free Calogero model, and 
\ba
\D^{(M)\pm}\eq \D^{(M)}\pm \de\D^{(M)}.\nn
\ea 
Then it follows from (\ref{comw}) that
\ba
[\HH,\DD^\pm]=\pm 2\w\DD^\pm\label{hdpm}
\ea
if we go from the Hamiltonian (\ref{HCOA}) to (\ref{tilhc}), following \cite{mag}, as in the free case.

The calculation of $\de \DD$ is completely similar to that of $\DD$ in the Appendix 4;
mainly, it amounts to using (\ref{AA}) again. Thus we present here only the result:
\ba
\de \DD=\de\LL-iN^{-1/2}\biggl[\de Q^+\phi_N+\phi_N^+\de Q^-\biggr]+
i\w N^{-1/2}y_N\biggl[2\phi_N^+\phi_N-\NN \biggr]\label{dedq}
\ea
where
\ba
\de Q^-\equiv w\sum_k x_k\psi_k\qq \de Q^+\equiv w\sum_k x_k\psi_k^+.\nn
\ea

Plugging (\ref{DDD}),(\ref{dedq}) into (\ref{ddpm}),we get:
\ba
\DD^\pm&=&\LL^{\pm}+iN^{-1/2}\biggl[(Q_f^+\mp\de Q^+)\phi_N-
\phi_N^+(Q_f^-\pm\de Q^-\biggr)+\nn\\
&+&(\NN -2\phi_N^+\phi_N)({\d\over\d y_N}\mp\w y_N)]\label{ddq}
\ea
where we mark the supercharges of the free Calogero model by the letter $f$. One can show that
the supercharges of the Calogero model in the oscillatory potential can be written as:
\ba
Q^\pm=Q^\pm_f+\de Q^\pm\qq \hat Q^\pm=Q^\pm_f-\de Q^\pm\nn
\ea
where $\hat Q^\pm $ are the supercharges with different sign of $\w$.

Then it follows from (\ref{ddq}) that
\ba
\DD^+&=&\LL^++iN^{-1/2}\biggl[\hat Q^+\phi_N-
\phi_N^+ Q^--(\NN -2\phi_N^+\phi_N)Q_N^+\biggr]\label{dd+}\\
\DD^-&=&\LL^-+iN^{-1/2}\biggl[ Q^+\phi_N-
\phi_N^+ \hat Q^-+(\NN -2\phi_N^+\phi_N)Q_N^-\biggr].\label{dd-}
\ea
As in the free case, it is helpful to separate the center of mass in the supercharges according to (\ref{Qqpm}).
In addition to (\ref{Qqpm}),(\ref{qccal}), one will then have for the quantities with inverted sign of $\w$,
\ba
\hat Q^\pm=\hat q^\pm+\hat Q_C^\pm\qq
\hat Q_C^-= -\phi_NQ_N^-\qq \hat Q_C^+= -\phi_N^+Q_N^+.\label{qtil}
\ea
Plugging (\ref{Qqpm}),(\ref{qccal}),(\ref{qtil}) into (\ref{dd+}), (\ref{dd-}), we get
\ba
\LL^+=\DD^+ +iN^{-1/2}\biggl[\phi_N^+q^--{\hat q}^+\phi_N-
\NN Q_N^+\biggr]\label{LL+}\\
\LL^-=\DD^- +iN^{-1/2}\biggl[\phi_N^+{\hat q}^--q^+\phi_N+
\NN Q_N^-\biggr].\label{LL-}
\ea

Now we are finally able to prove (\ref{HLLpm}) using (\ref{LL+}),
(\ref{LL-}) and (\ref{hdpm}).
We will consider only
the commutation with $\LL^+$ because the other one is just its
hermitean conjugation.

We will show that all the terms in the operator (\ref{LL+}) commute with
the superhamiltonian in accordance with (\ref{HLLpm}).
The first nontrivial commutator of that kind is:
\ba
[\HH,
\phi_N^+q^-]=[\HH,\phi_N^+]q^-=[2\w\NN,\phi_N^+]q^-=2\w\phi_N^+q^-.\nn
\ea

For the term containing ${\hat q}^+$ we need the superhamiltonian
(\ref{HHCO}) with $\w$ replaced by $-\w$:
\ba
{\hat \HH}=\HH-4\w\NN+2\w\biggl(1+(N-1)(Nl+1)\biggr)\qquad
[{\hat\HH},{\hat q}^\pm]=0.\nonumber
\ea
Then we can proceed with the commutators:
\ba
&&[\HH,{\hat q}^+\phi_N]=[{\hat \HH}+4\w\NN,{\hat q}^+\phi_N]=
[{\hat \HH},{\hat q}^+\phi_N]={\hat q}^+[{\hat \HH},\phi_N]={\hat
q}^+[-2\w\NN,\phi_N]=\nonumber\\&&=2\w{\hat
q}^+\phi_N.\nn
\ea
Finally,
\ba
[\HH,\NN
Q_N^+]=[\HH,\phi_N^+Q_C^-]=[\HH,\phi_N^+]Q_C^-=2\w\phi_N^+Q_C^-=2\w\NN
Q_N^+.\nn
\ea
If we now recall (\ref{hdpm}), we see that all the terms in the operator (\ref{LL+}) commute with
the superhamiltonian in accordance with (\ref{HLLpm}), so the latter is true.

Note that from (\ref{hdpm}) it follows that
\ba
[h,\DD^\pm]=\pm 2\w\DD^\pm\label{hDD}
\ea
where $h=\HH-H_C$ is the CM independent part of the Calogero
superhamiltonian (\ref{HHh}), where  $H_C$ is given by (\ref{hccal}).

Same as (\ref{HLLpm}), Eq. (\ref{hDD}) describes an oscillatory
algebra and hence can be used for the construction of the spectrum
of the superhamiltonian $h$ and proof of its integrability.
Namely, from (\ref{psiCO}) one can derive the ground state wave
function for $h$:
\ba
\psi_0=\exp\biggl[-{\w\over
2}\biggl(\sum_{j=1}^{N}x_j^2-y_N^2\biggr)\biggr]\prod_{i <
k}^N|x_i-x_j|^l.\nonumber
\ea
Applying powers of the operators 
(\ref{dd+}),(\ref{dd-})
 $q^\pm$ from (\ref{Qqpm}), and $\hat q^\pm$ from (\ref{qtil}) 
to this wave function one can get the excited states
of $h$, which parallels the construction from Subsection 2.3.

Eqs. (\ref{LL+}), (\ref{LL-}) are also useful for the derivation
of (\ref{TsL1}):
\ba
H^{(0)}=\Ts\L_1=\Ts\L_2+\const.\nonumber
\ea
The first equality of (\ref{TsL1}) can be proved in the following
way: taking into account (\ref{TsA}), we get
\ba
&&\Ts\L_1=\Ts(\L^+\L^-)=N<N|\LL^+\LL^-|N>= N<N|\biggl[
\DD^++iN^{-1/2}\biggl(\phi_N^+q^--\nonumber\\
&& -{\hat q}^+\phi_N-\NN Q_N^+\biggr)\biggr]\biggl[
\DD^-+iN^{-1/2}\biggl(\phi_N^+{\hat q}^--{ q}^+\phi_N- \NN
Q_N^-\biggr)\biggr]|N>=\nonumber\\
&&= N<N|iN^{-1/2}\biggl[\phi_N^+q^--
Q_N^+\biggr]iN^{-1/2}\biggl[-q^+\phi_N- Q_N^-\biggr]|N>=\nonumber\\
&&= -<N|-\phi_N^+q^-q^+\phi_N+Q_N^+Q_N^-|N>=<0|q^+q^-+Q_C^-Q_C^+
|0>=\nonumber\\
&&=<0|h+H_C|0>=H^{(0)}.\nonumber
\ea
The second equality of (\ref{TsL1}) can be proved in a similar
way:
\ba
&&\Ts\L_2=\Ts(\L^-\L^+)=N<N|\LL^-\LL^+|N>= N<N|\biggl[
\DD^-+iN^{-1/2}\biggl(\phi_N^+{\hat q}^--\nonumber\\
&&-{ q}^+\phi_N- \NN Q_N^-\biggr)\biggr]\biggl[
\DD^++iN^{-1/2}\biggl(\phi_N^+q^--{\hat q}^+\phi_N-\NN Q_N^+\biggr)\biggr]|N>=\nonumber\\
&&= N<N|iN^{-1/2}\biggl[\phi_N^+{\hat q}^--
Q_N^+\biggr]iN^{-1/2}\biggl[-{\hat q}^+\phi_N- Q_N^-\biggr]|N>=\nonumber\\
&&= -<N|-\phi_N^+{\hat q}^-{\hat q}^+\phi_N+Q_N^-Q_N^+|N>=
<0|{\hat q}^+{\hat q}^-+{\hat Q}_C^-{\hat Q}_C^+
|0>=\nonumber\\
&&=<0|{\hat h}+{\hat H}_C|0>={\hat
H}^{(0)}=H^{(0)}+2\w\biggl[1+(N-1)(Nl+1)\biggr]\nonumber
\ea
where the hat indicates the inversion of the sign of $\w$. 
We have used the fact that $\DD^\pm|N>=0;\ <N|\DD^\pm=0$, because $D^{(0)\pm}=0$.

We see
that (\ref{TsL1}) actually follows from the two supersymmetries of
the model.

\subsection*{\it 4.4.\quad The extension onto the root systems other
than $A_N$.}

In the papers \cite{Komori1}-\cite{Ujino3} Dunkl operators for the
root systems other than $A_N$ were introduced. The formalizm of the present text
can be extended to these more general models; in particular, one can define
analogs of the formulae (\ref{HA})-(\ref{HCOA}). For the
partial case (\ref{C}) of the Clebsh-Gordan coefficient, analogs
of the operators (\ref{DDD}), (\ref{ddpm}) that commute with the
superhamiltonian can be considered.

For the construction of the super Lax operators for general root
systems one should use the formalism of \cite{Lap} (bearing in mind
Appendix 1 from the present text).
Then it would be interesting to see the relation between 
the analogs of operators (\ref{LL}) and (\ref{DDD}) in this 
more general case (i.e., generalization of (\ref{DL}),(\ref{LD})).

\section*{\normalsize\bf Acknowledgements}

This work is a part of the author's Ph.D. thesis. The author is
grateful to the supervisor M. Ioffe for guidance and useful
discussions. The work has also been made possible in part by the
support provided by the grant of Russian Foundation of Basic
Researches N 02-01-00499.

\section*{\large\bf Appendix 1.}

 In the present Appendix we are to prove that for any
$V_{km}=-V_{mk}$ it is true that:
\ba
\sum_{m \neq k} V_{km}\psi_k^+\psi_k\KK_{km}&=&\sum_{m \neq k}
V_{km}\psi_k^+\psi_m.\label{VKD}
\ea

{\bf \ul{Proof}}:
\ba
\sum_{m \neq k} V_{km}\psi_k^+\psi_k\KK_{km}=\sum_{m \neq k}
V_{km}\psi_k^+\psi_k\biggl[\psi^+_k\psi_m+\psi^+_m\psi_k-\psi^+_k\psi_k-
\psi^+_m\psi_m+ 1\biggr] =\nonumber\\ =\sum_{m \neq k}
V_{km}\biggl[(1 -\psi_k\psi_k^+)\psi^+_k\psi_m+
\psi_k^+(\de_{km}-\psi_m^+\psi_k)\psi_k+(1-\psi_k\psi_k^+)\psi^+_k\psi_k
-\nonumber\\ -\psi_k^+ \psi_k\psi^+_m\psi_m+\psi^+_k\psi_k\biggr]
=\sum_{m \neq k} V_{km}\biggl[\psi^+_k\psi_m-\psi_k^+
\psi_k\psi^+_m\psi_m\biggr]=\sum_{m \neq k}
V_{km}\psi^+_k\psi_m\nonumber
\ea
because the contraction of a symmetric object $\psi_k^+
\psi_k\psi^+_m\psi_m$ and antisymmetric $V_{km}$ is zero.
\section*{\large\bf Appendix 2.}
  In this Appendix we shall prove the following
statement: for any $i,j$
\ba
(\widetilde T_{ij}^L)_{\a\xi} (\widetilde T_{ij}^{(M)})_{\ga
\b}<\z|C_\xi|\b>= <\nu| C_\a|\ga>(\widetilde
T_{ij}^{(M)})_{\nu\z}\label{TTC1}
\ea
where $\T_{ij}^L$ is defined in (\ref{til}); $\T_{ij}^{(M)}$ in
(\ref{defT}); $C_\xi$ in (\ref{C}).

We will need an auxiliary statement:
\ba
\KK_{ij}C_\b=C_\a(T_{ij}^L)_{\a\b}\KK_{ij}.\label{KC}
\ea
Proof of (\ref{KC}): it follows from (\ref{pe1}),(\ref{pr2}) that
\ba
\KK_{ij}\psi_i^+\psi_i=\psi_j^+\psi_j\KK_{ij}\qquad
\KK_{ij}\psi_k^+\psi_k=\psi_k^+\psi_k\KK_{ij}\qquad k\ne i,j\nonumber
\ea
(no summation over repeated indices). Hence, one can check that
\ba
\KK_{ij}\psi_k^+\psi_k=\psi_l^+\psi_l(T_{ij}^{(1)})_{lk}\KK_{ij}\nonumber
\ea
where the summation is only over $l$, and the matrix
$\T_{ij}^{(1)}$ is defined in (\ref{Tij1}). Then it follows that
\ba
 \KK_{ij}C_\b&=& \KK_{ij}R_{\b k}\psi_k^+\psi_k= R_{\b
k}(T_{ij}^{(1)})_{lk}\psi_{l}^+\psi_l \KK_{ij}= R_{\b
k}(T_{ij}^{(1)})_{lk}R_{ml}R_{mn}\psi_{n}^+\psi_n \KK_{ij} =\nonumber\\
&=& (T^L_{ij})_{m \b}C_m \KK_{ij}= (T^L_{ij})_{\mu \b}C_\mu \KK_{ij}
\nonumber
\ea
where we have used the identities:
\ba
R_{\b k}(T_{ij}^{(1)})_{lk}R_{ml}=(T^L_{ij})_{m \b}\qquad
 (T^L_{ij})_{N \b}=0\nonumber
\ea
proved in \cite{mag}.

Now we can use (\ref{KC}) to prove (\ref{TTC1}):
\ba
&&(T_{ij}^L)_{\a\xi} (T_{ij}^{(M)})_{\ga \b}<\z|C_\xi|\b>=
(T_{ij}^L)_{\a\xi}<\z|C_\xi|\b><\b|\KK_{ij}|\ga>=\nonumber\\ &=&
<\z|C_\xi (T_{ij}^L)_{\xi\a} \KK_{ij}|\ga> =<\z|\KK_{ij}
C_\a|\ga>=<\z|\KK_{ij}|\nu><\nu|C_\a|\ga>=\nonumber\\ &=&<\nu|
C_\a|\ga>(T_{ij}^{(M)})_{\nu\z}.\nonumber
\ea

\section*{\large\bf Appendix 3}

In the present Appendix we are to prove that

\ba
 D^{(M)}_{\s\a}&=&<\s|C_\xi|\b>R_{\k k}\biggl[-i\d_k\de_{\b\a} + i \sum_{m \neq k}
V_{km}({\widetilde T}_{km}^{(M)})_{\b\a}\biggr]=\nonumber\\
&=&<\s|\LL|\a> +iN^{-1/2}M{\d\over\d y_N}\de_{\s\a}\label{DL1}
\ea
where $\LL$ is the super Lax operator (\ref{LL}).

{\bf\ul{Proof}}:\\
One can modify the first line of (\ref{DL1}) in the following way:
\ba
 D^{(M)}_{\s\a}&=&<\s|C_\xi|\b>R_{\k k}\biggl[-i\d_k\de_{\b\a} + i \sum_{m \neq k}
V_{km}({\widetilde T}_{km}^{(M)})_{\b\a}\biggr]=\nonumber\\
&=&<\s|C_\xi|\b>R_{\xi
k}<\b|-i\d_k +i \sum_{m \neq k} V_{km}\KK_{km}|\a>=\nonumber\\
&=&<\s|C_\xi R_{\xi k}|\b><\b|-i\d_k +i \sum_{m \neq k}
V_{km}\KK_{km}|\a>.\label{DCR}
\ea

Taking into account the definition (\ref{C}) and the orthogonality
of $\R$, one can see that
\ba
C_\xi R_{\xi k}=R_{\xi k}R_{\xi l}\psi_l^+\psi_l=(\de_{kl}-R_{N
k}R_{Nl})\psi_l^+\psi_l=\psi_k^+\psi_k-N^{-1}\sum_{l}\psi_l^+\psi_l\label{CR}
\ea
where no summation over $k$ is implied. Thus,
\ba
<\s|C_\xi R_{\xi k}|\b> =<\s|\psi_k^+\psi_k -
N^{-1}\sum_{l}\psi_l^+\psi_l|\b>=<\s|\psi_k^+\psi_k -
N^{-1}M|\b>.\label{CRpsi}
\ea

Plugging (\ref{CRpsi}) into (\ref{DCR}), we get
\ba
&& D^{(M)}_{\s\a}=<\s|\psi_k^+\psi_k - N^{-1}M|\b><\b|-i\d_k + i
\sum_{m \neq k} V_{km}\KK_{km}|\a>=<\s|(\psi_k^+\psi_k
-\nonumber\\
&&-N^{-1}M)\biggl[-i\d_k  + i \sum_{m \neq k}
V_{km}\KK_{km}\biggr]|\a>=<\s|-i\psi_k^+\psi_k\d_k +i \sum_{m \neq
k} V_{km}\psi_k^+\psi_k\KK_{km} +\nonumber\\
&&+iN^{-1}M\sum_k\d_k- iN^{-1}M\sum_{m \neq k}
V_{km}\KK_{km}\biggr]|\a>.\label{DCK}
\ea
Plugging (\ref{VKD}) into (\ref{DCK}) and taking into account that
the contraction of a symmetric object $\KK_{km}$ and antisymmetric
$V_{km}$ is always zero, we get:
\ba
 D^{(M)}_{\s\a}&=&<\s|-i\psi_k^+\psi_k\d_k+i \sum_{m \neq k}
V_{km}\psi_k^+\psi_m +iN^{-1/2}M{\d\over\d y_N}
\biggr]|\a>=\nonumber\\ &=&<\s|\LL|\a> +iN^{-1/2}M{\d\over\d
y_N}\de_{\s\a}.\nonumber
\ea

\section*{\large\bf Appendix 4.}

In this Appendix we will determine the form of the operator (\ref{DDD}).
Plugging (\ref{DL}) into (\ref{DDD}), we get:
\ba
&&\DD=\sum_{M,\s,\a}<\s|\LL|\a>\biggl[|\s><\a|+|N\s><\a N|\biggr]+iN^{-1/2}{\d\over\d y_N}
 \sum_{M,\s}M\biggl[|\s><\s|+\nn\\
&&+|N\s><\s N|\biggr]\equiv\DD^{(1)}+\DD^{(2)}.\label{D1D2}
\ea

One could rewrite the operator $\DD^{(2)}$ as
\ba
\DD^{(2)}&=&iN^{-1/2}{\d\over\d y_N}\sum_{M,\s}M\biggl[|\s><\s|+|N\s><\s N|\biggr]=\nn\\
&=&iN^{-1/2}{\d\over\d y_N}\biggl(\SS +\phi^+_N\SS\phi_N\biggr)\label{DD20}
\ea
where
\ba
\SS\equiv\sum_{M,\s}M|\s><\s|=\sum_{M,\s,\b}\phi_\b^+\phi_\b|\s><\s|=
\sum_{\b}\phi_\b^+\phi_\b(1-\phi_N^+\phi_N).\label{MM}
\ea
It follows from (\ref{MM}) that
\ba
\phi^+_N\SS\phi_N=\phi^+_N\sum_{\b}\phi_\b^+\phi_\b(1-\phi_N^+\phi_N)\phi_N=
\phi^+_N\sum_{\b}\phi_\b^+\phi_\b\phi_N=\sum_{\b}\phi_\b^+\phi_\b\phi^+_N\phi_N.\label{MM2}
\ea
Plugging (\ref{MM}) and (\ref{MM2}) into (\ref{DD20}), one obtains:
\ba
\DD^{(2)}=iN^{-1/2}{\d\over\d y_N}\sum_{\b}\phi_\b^+\phi_\b.\label{DD2}
\ea

To get an explicit form of $\DD^{(1)}$ in (\ref{D1D2}), note: for any operator $\AA$ of the form (\ref{AAbil}),
\ba
&&\sum_{M,\s,\a}<\s|\AA|\a>\biggl[|\s><\a|+|N\s><\a N|\biggr]=
\AA-N^{-1/2}\sum_{k,m}A_{km}\biggl[\psi_k^+\phi_N+\nn\\
&&+\phi_N^+\psi_m\biggr]+N^{-1}\phi_N^+\phi_N\sum_{k,m}A_{km}\label{AA}
\ea
where $\NN|\s>=M|\s>; \NN|\a>=M|\a>$.
The proof of this relation is rather long and we will not give it.
In short, it uses the following auxillary relation:
\ba
\sum_{M,\s,\a}<\s|\AA|\a>\biggl[|\s><\a|&+&|N\s><\a N|\biggr]
=\AA+[\phi_N^+,\AA]\phi_N-\phi_N^+[\phi_N,\AA]+\nn\\
&+&\{\phi_N^+,[\phi_N,\AA]\}\phi_N^+\phi_N\nn
\ea
that is true for {\it any} fermionic operator $\AA$, not necessarily bilinear, and follows from
the completeness of the basis (\ref{ste}) and anticommutation relations (\ref{phiant}).

Taking into account (\ref{AA}) for the operator $\LL$, we get:
\ba
\DD^{(1)}=\LL-N^{-1/2}\sum_{k,m}L_{km}\biggl[\psi_k^+\phi_N+
\phi_N^+\psi_m\biggr]+ N^{-1}\phi_N^+\phi_N\sum_{k,m}L_{km}.\label{D1LL}
\ea

Since the super Lax operator $\LL$ has the form (\ref{LL}), one can check that \cite{Shastry}
\ba
\sum_{m}L_{km}=-iQ^-_k\qq
\sum_{k}L_{km}=iQ^+_m\qq
\sum_{k,m}L_{km} =-iN^{1/2}{\d\over\d y_N}\label{Lkm}
\ea
where the operators $Q_i^\pm$ are defined by (\ref{Qf}) and $y_N$ by (\ref{jac}). Therefore, 
\ba
\sum_{km}L_{km}\psi_k^+=-iQ^+\qq
\sum_{km}L_{km}\psi_m=iQ^-\label{LQ}
\ea
where $Q^\pm$ are supercharges (\ref{Qpm}) with bosonic part (\ref{bocal}).

Plugging (\ref{Lkm}) and (\ref{LQ}) into (\ref{D1LL}) we get:
\ba
\DD^{(1)}=\LL+iN^{-1/2}\biggl[Q^+\phi_N-\phi_N^+Q^--\phi_N^+\phi_N{\d\over\d y_N}\biggr].\label{DD1fin}
\ea
Bringing together (\ref{DD1fin}), (\ref{DD2}) and (\ref{D1D2}) one sees that
\ba
\DD&=&\LL+iN^{-1/2}\biggl[Q^+\phi_N-\phi_N^+Q^-+\biggl(\sum_\b\phi_\b^+\phi_\b-
\phi_N^+\phi_N\biggr){\d\over\d y_N}\biggr]=\LL+\nn\\
&+&  iN^{-1/2}\biggl[Q^+\phi_N-\phi_N^+Q^-+\biggl(\NN-
2\phi_N^+\phi_N\biggr){\d\over\d y_N}\biggr].\nn
\ea



\vspace{.5cm}
\section*{\normalsize\bf References}
\begin{enumerate}
\bibitem{Perelomov}
   Olshanetsky M A and Perelomov A M 1983 {\it Phys. Rep.} {\bf 94} 6
\bibitem{genlax}  Calogero F 1975
{\it Lett. Nuov. Cim. } {\bf 13} 411
\bibitem{Calogero1}  Calogero F 1971 {\it Journ. Math. Phys.} {\bf 12} 419

\bibitem{Vasiliev0}
   Brink L Hansson T H and Vasiliev M A 1992 {\it Phys. Lett.} B {\bf 286} 109


\bibitem{Ruhl}
Ruhl W and Turbiner A 1995 {\it Mod. Phys. Lett.} A {\bf 10} 2213
\bibitem{Moser}
Moser J 1975 {\it Adv. Math.} {\bf 16} 197


\bibitem{WU}
Ujino H and Wadati M 1997 {\it J. Phys. Soc. Japan} {\bf 66} 345

\bibitem{sinh}  Calogero F Ragnisco O and Marchioro C
1975 {\it Lett. Nuov. Cim.} {\bf 13} 383

 \bibitem{Suther1}
    Sutherland B 1971 {\it Phys. Rev.} A {\bf 4} 2019
\bibitem{Suther2}
Sutherland B 1972 {\it Phys. Rev.} A {\bf 5} 1372

\bibitem{Lap0}
 Lapointe L  Vinet L 1996 {\it Comm. Math. Phys.} {\bf 178}425


\bibitem{Sasakibig}
Khastgir S P Pocklington A J and Sasaki R 2000 {\it J. Phys. A:
Math. Gen.} {\bf 33} 9033
\bibitem{Komori1}
Komori Y 1998 {\it Lett. Math. Phys.} {\bf 46} 147
\bibitem{Komori2}
Komori Y 2000 {\it Physical Combinatorics} ed M Kashivara and T
Miwa (Boston: Birkh\"auser) p141
\bibitem{Ujino1}
Nishino A Ujino H Komori Y and Wadati M 2000 {\it Nucl. Phys.}
B {\bf 571} 632
\bibitem{Ujino2}
Nishino A and Wadati M 2000 {\it J. Phys. A: Math. Gen.} {\bf 33}
3795
\bibitem{Ujino3}
Nishino A and Ujino H 2001 {\it J. Phys. A: Math. Gen.} {\bf 33}
4733
\bibitem{Turbiner}
    Brink L  Turbiner A  and Wyllard N 1998 {\it  J.  Math.  Phys.}
{\bf 39} 1285
\bibitem{UHW}
  Ujino H Hikami K and Wadati M 1992 {\it J. Phys. Soc. Jpn.} {\bf
  61} 3425
\bibitem{UWH}
Ujino H  Wadati M and Hikami K  1993 {\it J. Phys. Soc. Jpn.} {\bf
  62} 3035
\bibitem{WU0}
Ujino H and Wadati M 1996 {\it J. Phys. Soc. Japan} {\bf 65} 2423
\bibitem{BMS}
Bordner A J Manton N S and Sasaki R 2000 {\it Prog. Theor. Phys.}
{\bf 103} 463.

\bibitem{Rittenberg}
de Crombrugghe M and Rittenberg V 1983 {\it Ann. Phys.} {\bf 151}
99
\bibitem{abi}
 Andrianov A  A  Borisov N  V  Ioffe M  V  and Eides M I 1984 {\it Phys. Lett.
A: Math. Gen.} {\bf 109} 143  \\
 Andrianov A  A  Borisov N  V  Ioffe M  V  and Eides M I  1985
{\it Theor. Math. Phys.} {\bf 61} 965 [transl  from 1984 {\it
Teor. Mat. Fiz.}
{\bf 61} 17] \\
 Andrianov A  A  Borisov N  V  and Ioffe M  V 1984 {\it Phys. Lett.} A {\bf
105} 19  \\
 Andrianov A  A  Borisov N  V  and Ioffe M  V 1985 {\it Theor. Math. Phys.}
{\bf 61} 1078  [transl  from 1984 {\it Teor. Mat. Fiz.} {\bf 61}
183]
\bibitem{Freedman}
   Freedman D  Z  and Mende P  F 1990 {\it Nucl.  Phys.}
B {\bf 344} 317
\bibitem{Vasiliev1}
    Brink L  Hansson T  H  Konstein S  E  and Vasiliev M  A 1993 {\it
Nucl.  Phys.} B {\bf 401} 591
\bibitem{Shastry} Shastry B S  and Sutherland B 1993 {\it Phys.  Rev.  Lett.} {\bf
70} 4029
\bibitem{Lap}
 Desrosiers P  Lapointe L and Mathieu P 2001 {\it Nucl. Phys.} B  {\bf 606} 547

\bibitem{dunkl}
   Dunkl C  F 1989 {\it Trans.  Amer.  Math.  Soc.}
 {\bf 311} 167
 \bibitem{poly1}
Polychronakos A  P 1992 {\it Phys.  Rev.  Lett.}
 {\bf 69} 703
\bibitem{poly2}
 Minahan J  A  and Polychronakos A  P  1993 {\it Phys  Lett} B {\bf
302} 265
\bibitem{Pas}
Pasquier V hep-th/9405104
\bibitem{Ghosh}
 Ghosh P  Khare A  and Sivakumar M  1998 {\it Phys.  Rev.} A {\bf 58} 821
\bibitem{Sasin}
Inozemtsev V I and Sasaki R hep-th/0105164
\bibitem{Des2}
Desrosiers P Lapointe L and Mathieu P 2001 {\it Nucl. Phys.} B
{\bf 606} 547
\bibitem{Des3}
Desrosiers P Lapointe L and Mathieu P hep-th/0305038.
\bibitem{inter}
 Ioffe M V and Neelov A I 2002 {\it J. Phys. A: Math. Gen.} {\bf 35} 7613
\bibitem{mag}
 Ioffe M  V  and Neelov A  I 2000 {\it J.  Phys. A: Math. Gen.} {\bf 33} 1581
\bibitem{Reed}
 Reed M  and Simon B 1978 {\it Methods of modern mathematical physics}
vol  III (New York: Academic)
\bibitem{eft}
  Efthimiou C  and Spector H 1997 {\it Phys. Rev.} A
 {\bf 56} 208

\bibitem{Ham}
 Hamermesh M 1964 {\it Group Theory and its application to physical problems}
(New York: Addison-Wesley)










\end{enumerate}
\end{document}